# Simulating Urban Scaling with an Term Linkages Network of a University

Anthony F.J. van Raan

Centre for Science and Technology Studies, Leiden University, Leiden, The Netherlands
vanraan@cwts.leidenuniv.nl
ORCID: 0000-0001-8980-5937

*Abstract*

*In this paper we make an attempt to increase our understanding of the urban scaling phenomenon. The aim is to investigate how superlinear scaling emerges if a network increases in size and how this scaling depends on the occurrence of elements that constitute the network. To this end we consider a city as a complex network structure and simulate this structure by the network of all publications of a research intensive university. In this simulation the publications take the role of the city inhabitants and the concepts (terms, keywords) in the publications represent all kinds of abilities and qualities of the inhabitants. We use in this experiment all author- and database-given terms of the scientific publications of Leiden University from 2022. We calculate the co-occurrence of terms, and on the basis of these connections we create a network and let this network grow by successively adding publications from the total set of publications. In this way we get a series of networks with different sizes and this simulates a series of cities with different number of inhabitants. This procedure is performed for different values of the term occurrence threshold. We then analyze how four important network parameters, namely number of terms, number of clusters, number of links, and total link strength increase with increasing size of the network. Particularly the number of network links and the total network linkage strength are in our opinion the parameters that dominate the scaling phenomenon and can be considered as a simulation of the socio-economic strength of a city, i.e., its gross urban product. We find a significant power law dependence of these network parameters on network size and the power law exponents for the lowest occurrence threshold are within the range that is typical for urban scaling. In our approach the number of clusters can be interpreted as a measure of complexity within the network. Since the occurrence threshold determines the diversity of terms, we may expect a special relation between the occurrence threshold and the number of clusters. This is indeed the case: whereas for the three other network parameters the scaling exponent increases with increasing occurrence threshold, the number of clusters is the only network parameter of which the scaling exponent decreases with increasing occurrence threshold. Finally, we discuss how our publication term network approach relates to scaling phenomena in cities.*

# 1. Introduction

Recent studies show a *more than proportional* (superlinear) increase of the socioeconomic performance of cities (measured by the gross urban product) in relation to population size (Bettencourt et al., 2007; Bettencourt et al., 2010; Lobo, Bettencourt et al., 2013; Bettencourt and Lobo, 2016; Cebrat and Sobczyński, 2016; Barthelemy, 2019a,b). If *individual cities i* with population $P_i$ have a gross urban product $G_i(P_i)$ then we find for a *set of cities*, for example all cities within a country, the relation

$$G(P) = AP^b \quad (1)$$

This *urban scaling* relation implies a power-law dependence of the gross urban product with population size. The coefficient $A$ and the exponent $b$ follow from the measurement; in most cases, values of the exponent are between 1.05 and 1.25. We refer to our recent work on urban scaling for further details (van Raan, 2016; 2020). As an example we give in Figure 1 the urban scaling of the major cities in the Netherlands.

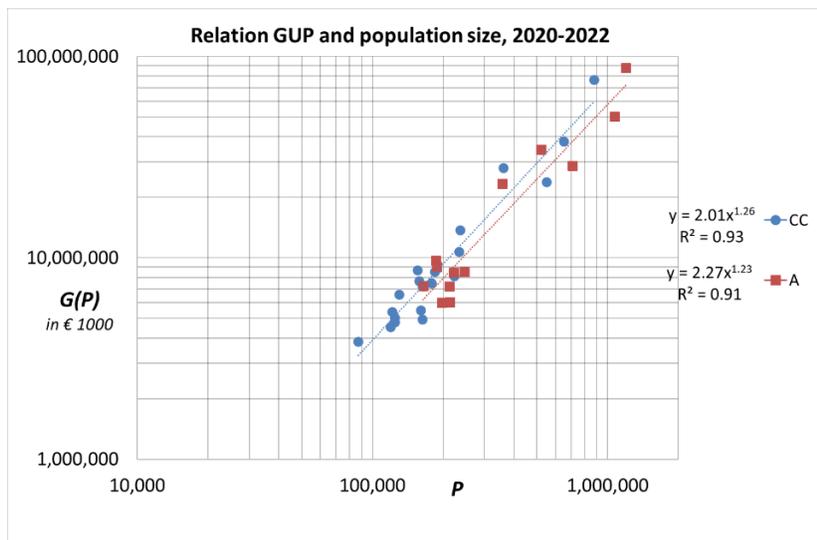

*Figure 1. Scaling of the gross urban product G(P) with population size P for the major cities in the Netherlands. The central cities are indicated with blue diamonds (C), their urban agglomerations with red squares (A). Data are the average for the years 2020-2022.*

Urban scaling behavior is also found for other city characteristics such as employment in terms of number of jobs, number of crimes, total road length and knowledge production activities in cities (Bettencourt et al., 2007; Arbesman et al., 2009; Alves et al., 2013; Schläpfer et al., 2014).

A simple way to understand the phenomenon of urban scaling is by seeing cities as a complex network. The larger the city in population size, the more network nodes. The nodes in the urban system are abilities and capacities of the inhabitants, social and cultural institutions, centers of education and research, firms, etcetera. The number of nodes has a *linear* dependence on size, but the links between nodes depend on size in a *superlinear* way.

The links between these (clustered) nodes represent many types of exchange activities such as movement of people, transportation of goods, transmission of information (Shutters et al., 2018), exchange of ideas and expertise. Therefore, these links are crucial for new developments, reinforcement of urban facilities, and innovation. One can expect that the productivity and efficiency of an urban economy increases as the level of connectivity among the constituent units increases. Because they increase superlinearly, the socioeconomic strength of cities increases more than proportionally with increasing population size. In this way, cities can be conceived as a complex network structure (Qian et al., 2020; Stier et al., 2022) following scaling laws (Batty, 2008; 2009).

The measurement of urban scaling is complicated by a number of problems. Scaling characteristics may depend on the definition of cities in terms of population size and population density of the urban area, and even on the size of a city measured in terms of residents and of working population size (Strumsky et al., 2021). These different definitions of cities can lead to finding linear scaling in situations where other researchers find superlinear scaling (Arcaute et al., 2015; Cottineau et al., 2017). The scaling exponent can even change from superlinear to sublinear (Louf and Barthelemy, 2014a). The chosen set of cities may also lead to different scaling exponents (Cebrat and Sobczyński, 2016; van Raan, 2020; Sahasranaman and Bettencourt, 2021). For cities in the richer western part of the EU a lower scaling exponent was found as compared to cities in the less rich eastern part of the EU (Strano and Sood, 2016). Scaling exponents may also vary in time (Cebrat and Sobczyński, 2016; Strano and Sood, 2016; Depersin and Bathelemey, 2018; Bettencourt et al. 2020; Strumsky et al., 2021). Scaling exponents differ for the various variables such as for instance gross domestic product, employment and number of patents (Bettencourt and Lobo, 2016). Ramaswami et al (2018) show that the economic structure of cities may considerably affect the measured urban scaling exponents. An Australian study (Sarkar et al., 2018; Sarkar, 2019) shows that scaling exponents may also differ within a broader variable: for lower income the scaling is about linear but for high income the scaling is superlinear. Leitão et al. (2016) discuss the problems in the statistical estimation of scaling exponents. Nevertheless, an extensive study on urban scaling in Brazil (Meirelles et al., 2018) using a large number of variables and different statistical approaches shows that socioeconomic variables (e.g., GUP) scale superlinearly with city population size, whereas infrastructural variables (e.g., length of street network) scale sublinearly and individual basic services (e.g., number of houses) scale linearly.

In understanding urban scaling, it is generally assumed that the scaling of urban economic performance is driven by social and economic interactions that increase disproportionally with urban population size. Ribeiro et al. (2017) for instance discuss a model of urban scaling based on the distant-dependent interaction range between the citizens and on the spatial structure of a city. Altmann (2020) studies interactions between individuals in different cities and shows that including these inter-city interactions leads to better scaling models.

There are several important studies on the theoretical derivation of urban scaling based on mathematical frameworks (Bathelemy, 2019a,b; Bettencourt, 2019), often related to statistical physics. Lobo et al. (2020) use a theoretical settlement model to explain urban scaling. The growth of cities is studied, for instance, by a diffusion-limited aggregation model (Makse et al., 1995) and by a population dynamics model (Reia et al., 2022). A theoretical framework of local interactions was used to study scaling properties of cities (Bettencourt and West, 2010; Bettencourt, 2013). Our approach is not a theoretical, mathematical construction, but an experiment with a 'real-world' growing network of which the parameters are measured as a function of network size. Thus our study poses a specific question about scaling, namely can we discover scaling properties similar to urban scaling if we analyze the linkage structure of a growing network and can we find how this scaling depends on the occurrence of elements that constitute the network. We think that this is a novel approach in research on urban scaling.

There are many studies on the properties of network structures, cluster formation, community detection and connectivity of nodes (Barabasi and Albert, 1999; Jin et al., 2001; Newman, 2004; Newman and Girvan, 2004; Newman, 2006; Traag et al., 2011; Fortunato and Newman, 2022). Network approaches have been applied to study how cities evolve (Batty, 2008; 2013), and for instance Lenormand et al. (2015) use an network approach to develop clustering methods for the study of scaling in land use organization in cities. However, there is little research on the enlargement of real-world networks (i.e., constructed with elements from existing systems as opposed to theoretical constructions) and its relation with connectivity (number of links, total link strength) and clustering (number of clusters), and, in particular, the scaling of these network parameters as a function of network size. Thus, to the best of our knowledge there are no studies that relate urban scaling phenomena to scaling properties of growing networks.

As we stated earlier, urban scaling is fundamentally based on social interactions. In this approach a city is a network of social interactions and the total number of interactions is directly related to the socio-economic output of cities. This directly relates to research on economic models of agglomeration effects that are based on an increasing productivity derived from social interactions (Glaeser et al., 1995; Meijers et al., 2016; Strumsky et al. 2023). Cottineau et al. (2019) analyse urban clusters to detect agglomeration economies.

The structure of this paper is as follows. Section 2 addresses our method and data, and in particular why we simulate a city with a university. In Section 3 we present and discuss our results. We conclude this paper in Section 4 with a discussion of the implications for urban scaling.

## 2. Method and Data

### *2.1 Simulating a city with a university*

In an earlier study (van Raan 2013) we found that also universities show a scaling behavior similar to cities. Whereas for cities the gross urban product is an important measure of socio-economic strength, for universities the extent to which all publications of a university are cited worldwide is an important measure of the scientific strength of the university. It is this analogy that we use to simulate the urban network in order to find and analyze scaling properties. Our aim is to investigate how superlinear scaling emerges if a network increases in size, and how this scaling depends on the occurrence of elements that constitute the network. To this end we consider a city as a complex network structure and simulate this structure by the network of all publications in a research intensive university. In this simulation the publications take the role of the city inhabitants and the concepts (terms, keywords) in the publications represent all kinds of abilities and qualities of the inhabitants. We then assume that the total links strength of the publication terms network is a analogon of the socioeconomic performance of a city. In this way we investigate scaling of growing networks in an empirical approach.

We use in this experiment all author- and database-given keywords of the scientific publications of Leiden University from 2022. Using publications of a large university means that one covers all fields of science, and this also resembles a city: a very heterogenous content within one coherent system. Just like a city, a university is a not an isolated system but a system of which the content is strongly related with the outside world.

We calculate the co-occurrence of terms, and on the basis of these connections we create a network and let this network grow by successively adding publications from the total set of publications. In this way we get a series of networks with different, increasing sizes and this simulates a series of cities with different number of inhabitants. This procedure is performed for different values of the term occurrence threshold. We then analyze how four important network parameters, namely number of terms, number of clusters, number of links, and total link strength increase with increasing size of the network. Particularly the number of network links and the total network linkage strength are the parameters that dominate the scaling phenomenon and can be considered as a simulation of the socio-economic strength of a city, i.e., its gross urban product.

Shutters et al. (2018) find that the increase in network density is for an important part driven by the presence of rare, but highly interdependent occupations in a city. This supports our approach to particularly include rare terms (i.e., by using low occurrence thresholds, see next section) in the network construction. For examples of bibliometric (e.g., co-citation, term co-occurrence) networks we refer to Price (1965); Braam et al., 1991; Lehmann et al 2003; Ren et al. (2012); Martin et al., 2013; Golosovsky and Solomon (2017); Muppidi and Reddy (2020). An overview of bibliometric network analysis is given by Perianes-Rodriguez et al. (2016).

A particularly interesting paper for our study is the work on the spatial structure of mobility networks in cities (Louail et al., 2014; 2015; Louf and Barthelemy, 2013; 2014b; 2016). These authors find that the number of hotspots (urban activity centers) scales sublinearly with city population size and they remark that this finding may serve as a guide for constructing a theoretical model. As we will see further on in this paper, we find a similar scaling behavior in our simulation network.

## 2.2 Creating a network

Given that publications can be characterized by a number of terms, a matrix of publication-to-term relations can be constructed. Using linear algebra, from this matrix a new matrix of term-to-term relations can be deduced (van Raan, 2019). The off-diagonal elements of this term-to-term (term co-occurrence) matrix indicate the number of publications in which the two terms involved are both present, i.e., the *co-occurrence frequency* of this pair of terms. The larger the number of publications in which two terms co-occur, the stronger the terms are considered to be related to each other, i.e., the stronger their link. The network grows with the number of publications, but the network itself is a structure of interconnected terms. This is similar to what we want to simulate: the size of cities is determined by the number of inhabitants, whereas the real basis of the urban scaling phenomenon is the network of interactions between all possible abilities and qualities of these inhabitants. The size of a city therefore is a kind of an easy measurable, proxy independent variable.

In order to construct and to visualize this bibliometric network we apply the software tool VOSviewer (van Eck and Waltman, 2010). Publications can be uploaded from large publication databases -such as the Web of Science or Scopus- into the VOSviewer system on the basis of their full database records including author- and database-given terms (keywords). The VOSviewer algorithm calculates how many co-occurrence links there are in the uploaded set of publications, the strength of the links (number of publications in which two terms occur together) and the total link strength. The terms and their links constitute the bibliometric network. In addition, terms with a high link strength are grouped into clusters. The matrix of term co-occurrence frequencies serves as input for the VOSviewer mapping technique. This technique determines for each term a location in a two-dimensional space. Strongly related terms tend to be located close to each other in the two-dimensional space, whereas terms that do not have a strong relation are located further away from each other. For brevity we will use in the remainder of this paper 'term map' instead of term-occurrence map. The VOSviewer mapping technique is related to multidimensional scaling, but for the purpose of creating term maps it has been shown to yield more satisfactory results. The VOSviewer clustering technique is a modularity-based community detection procedure which is discussed in detail by Van Eck et al. (2010), Waltman et al. (2010), Waltman and Van Eck (2013).

It is possible to set a threshold value for the *minimum number of term occurrences*. We can use this threshold parameter to specify the minimum number of occurrences that a term must have within the set of publications to be included in a term map. We experiment with different *occurrence thresholds*, beginning with threshold value $\gamma=1$, i.e., *all* terms identified in the set of publications are taken into account in the construction of the network, and further threshold values up to $\gamma=10$. Thus, the occurrence threshold determines the *diversity of terms*: at a low threshold most or all of the terms are included in the network formation, whereas at a high threshold only the more frequently occurring terms are included.

As mentioned above, the VOSviewer methodology also includes a technique in which terms with relatively high co-occurrence links are grouped into a cluster. What is the meaning of these clusters in the context of our simulation of cities? As discussed previously, the terms in the network correspond to the abilities and capacities of the inhabitants and the co-occurrences are the connections between these abilities and capacities. This will give rise to organizations and institutions in the educational, research, business, cultural and social sectors, which become visible as clusters in the network. In this sense, the number of clusters is a measure of complexity within the network. We may expect a relation between diversity (value of occurrence threshold) and complexity (number of clusters). In Section 3 we will see that our analyses indeed show a relation between the occurrence threshold and the number of clusters.

## *2.3 Network Data and Network Parameters*

We collect from the Web of Science (WoS) all 7,653 publications of Leiden University published in 2022. Leiden University is a research intensive university, oldest in the Netherlands (founded 1575), 34,000 students, invariably in the top-100 worldwide of the international rankings, 15 Nobel Laureates affiliated to the university (Leiden University 2023). The full records of the Leiden publications including author- and database-given terms (keywords) are uploaded in the VOSviewer. All 7,653 publications are put in alphabetic order on the basis of the name of the first author. We assume that this assures a random composition of the set of publications as far as the subjects of research (and with that the relevant terms) concerns. The occurrence frequency of terms may differ substantially for the various terms. Some terms occur only in one or a few publications (these publications relate to a very specific topic such as supermassive black holes), and other terms occur in many publications (often these publications deal with important ad general topics in medical research, for instance cancer or quality-of-life).

A representative sample of 7,653 is 366 (confidence level 95% and margin of error 5%) so we start with alphabetically the first 400 Leiden publications and successively double the number: 800, 1,600, 3,200 and so on. As a check we also take the first 1,000, and then 2,000, 3,000 up to 7,000 publications. In this way, we let the network grow: we have enlarged the network structure from 400 to 7,000, an enlargement of nearly a factor 20.

To give an impression of the visualization of the Leiden publications by the VoS viewer mapping procedure we show in Figure 2 the term map representing the network for the (alphabetically) first 6,000 publications with occurrence threshold $\gamma=4$ (we take a higher occurrence threshold in order to avoid overloading the map with a very large amount of terms, we now have in total 2,222 terms). We see a large medical research cluster (left side) and an astronomy and astrophysics cluster (right side). In the network maps the different clusters are indicated with different colors, the size of the circle of a term is determined by its weight, i.e., the number of publications in which this term occurs. For details of the network and mapping procedures we refer to the VOSviewer manual (van Eck and Waltman 2020).

*Figure 2. Term map of Leiden University publications (2022), n=6000, γ=4, the map is constructed with all 2,222 terms.*

## 3. Results and Discussion

### 3.1 Network parameters at different occurrence thresholds for a growing network

For each network structure the VoSviewer calculates four parameters: the number of terms involved in the construction of the network, the number of clusters, the number of links, and the total strength of the links. These are our network parameters of which we want to find the dependence on the size of the network. We perform these calculations for occurrence thresholds γ=1 to 10.

In Figure 3 we show our results for occurrence thresholds γ=1, 2, and 3; the results for γ= 1 up to 10 are presented in the Supplementary Material Table S1. We find for our four network parameters $p_i$ (total link strength $p_1$, number of links $p_2$, number of terms $p_3$, number of clusters $p_4$) a significant power law dependence on network size, so we can write

$$p_{i,\gamma}(n) = B(i,\gamma).n^{\beta(i,\gamma)} \qquad (2)$$

Where $p_{i,\gamma}$ is network parameter *i* at term occurrence threshold $\gamma$, $n$ is the size of the network, $B(i,\gamma)$ is the empirically determined equation coefficient for network parameter $p_{i,\gamma}$ at term occurrence threshold $\gamma$, and $\beta(i,\gamma)$ is the power law exponent empirically determined for network parameter $p_{i,\gamma}$ at term occurrence threshold $\gamma$. We find in all cases $0 \leq \beta(i,\gamma) \leq 2.5$. We see that for the lowest occurrence threshold $\gamma=1$ the total link strength as well as the number of links increase slightly *superlinearly*, for the higher occurrence thresholds the increase is strongly *superlinear*. The number of terms increases for $\gamma=1$ *sublinearly* with network size, but for the higher occurrence thresholds *superlinearly*. The number of clusters is the only parameter which, regardless of the value of the occurrence threshold, increases sublinearly. We also see that the scaling coefficient $B(i,\gamma)$ which determines the absolute value of the network parameters, decreases with increasing occurrence threshold for all parameters.

We also performed our calculations after removing those terms that have no or a very small link strength (link strength lower than 10; this value is reasonable given the link strength distribution of all terms in the network, see Supplementary Material Figures S1 and S2). For example, in the case of network size 6,000 and $\gamma=1$, 27% of the terms is removed and that leads to 13% less total link strength of the entire network. Generally, for most of the network parameters the measured scaling exponents do not change much (see Supplementary Material Table S1). The scaling exponent for $\gamma=1$ increases from 1.04 to 1.08. However, what changes drastically by removing terms with no or a very small link strength is the number of clusters. In the case of network size 6,000 and $\gamma=1$, the number of clusters decreases from 245 to 88: removing terms with no or a very small link strength means that many small clusters disappear.

In Section 2.2 we discussed that the number of clusters can be seen as a measure of complexity and that a specific relation with the occurrence threshold can be expected. We see in Figure 3 this specific relation: whereas for the three other network parameters the scaling exponent increases with increasing occurrence threshold, the number of clusters is the only network parameter of which the scaling exponent decreases with increasing occurrence threshold. Interestingly, the number of clusters increases *sublinearly* with network size particularly for the lower occurrence thresholds. This finding is interesting because the network clusters can be seen as activity centers and other studies show a similar sublinear increase of activity centres ('hotspots') in cities (Louail et al., 2014; 2015; Louf and Bathelemy, 2013; 2014b; 2016). Particularly with occurrence thresholds above 2, the number of clusters hardly increases with increasing size. This suggests that in a growing network the complexity does increase but not 'explode' and this in turn may be an indication that a growing network tend to decrease complexity, which may be a sign of effective self-organization. Further research, however, is necessary to rule out effects of the VOSviewer clustering algorithm at a higher number of publications.

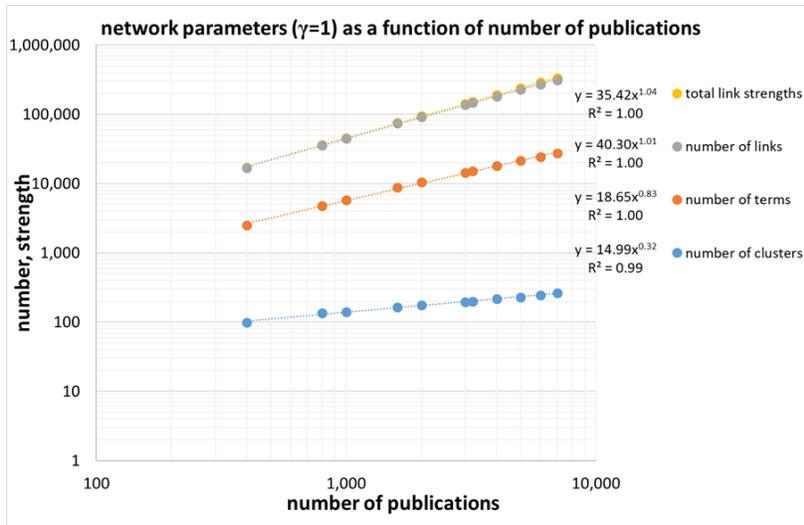
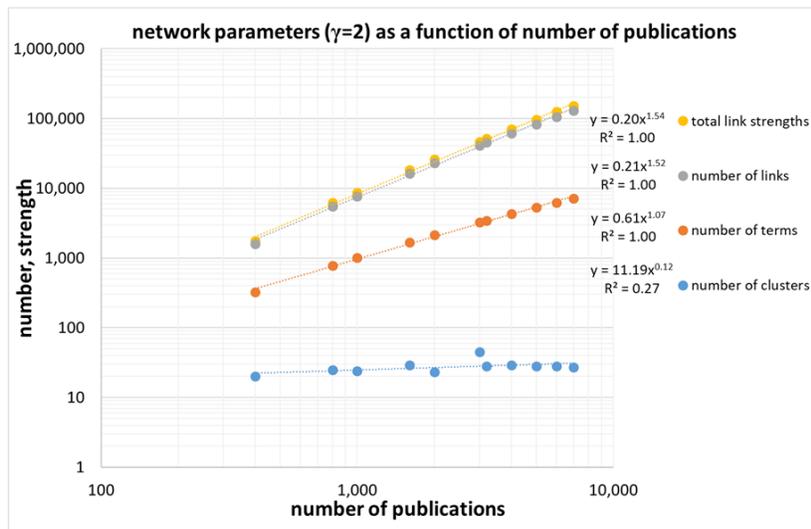
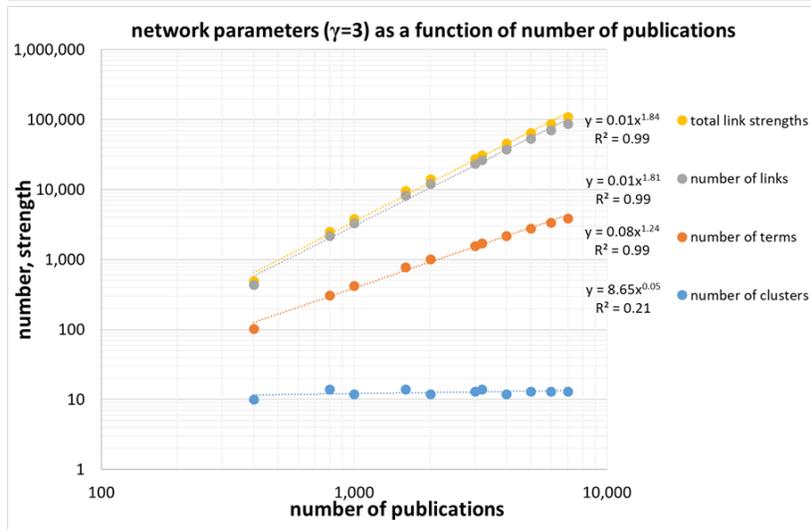

*Figure 3. Network parameters as function of network size (number of publications) and occurrence thresholds $\gamma=1$ (upper panel), 2 (middle panel), and 3 (lower panel).*

To the best of our knowledge, the only empirical work on the increase of the number of links for a growing network is the study of Li and Lu (2009) and Li (2009). These authors find values between 1.12 and 1.31 which is close to our values, see Supplementary Material Section 3.

### *3.2 Influence of term occurrence and of occurrence threshold*

Our measurements on the growing network as presented in Figure 3 are performed with specific occurrence *thresholds* $\gamma$, for instance $\gamma=1$ means all terms with occurrence 1 and higher are included. It is interesting to find out how specific occurrence values play a role in the structure of the network. In the Supplementary Material Figure S3 we present the relation between the number of terms and the occurrence of a term, and the relation sbetween the occurrence of a term and its link strength in the network. We see that the number of terms decreases as a function of occurrence with a power law exponent -2.33. On the other hand, the higher the occurrence of a term, the larger its link strength in the network, with a linear relation. Thus, the higher the occurrence of a term, the lower the number of these terms, but the larger its contribution to the total link strength of the network.

In the previous section we discussed the dependence of the network parameters on network size for different occurrence thresholds. We now analyze the data as a function of the occurrence threshold. With the data used in Figure 3 we can analyze the relation between the scaling exponents and the occurrence threshold more explicitly. The results are presented in Figure 4. We find that the higher the term occurrence threshold, i.e., the more common terms are involved, the higher the power law exponents, except for the number of clusters.

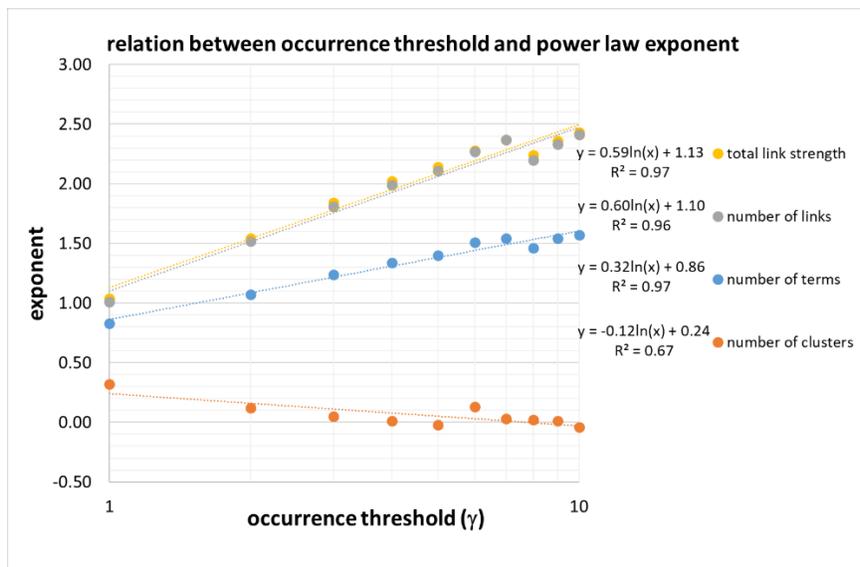

*Figure 4. Relation between occurrence threshold and power law exponents of the network parameters.*

Figure 4 shows the relatively large difference between the scaling exponent for $\gamma=1$ and the scaling exponents for the higher occurrence thresholds. This has an interesting consequence. In the case of small networks, the total link strength is the largest for $\gamma=1$ (highest term diversity). But if the network becomes larger then, notwithstanding its small scaling coefficient, the total link strength for $\gamma=5$ (which means a lower term diversity) will exceed the total link strength for $\gamma=1$ because of its high scaling exponent. It can be easily shown by equating the scaling equations for $\gamma=1$ and 5 that this will happen around $n=25,000$. This would imply that for very large networks term diversity plays a declining role in the total link strength.

With the results shown in Figure 4 we can analyze the relation between the scaling exponents and the occurrence threshold more explicitly. We find that the higher the term occurrence threshold, i.e., the more common terms are involved, the higher the power law exponents, except for the number of clusters, and this relation is with high significance logarithmic:

$$\beta(i,\gamma) = a(i)ln\gamma + b(i) \qquad (3)$$

where the coefficient $a(i)$ is positive for the number of terms, the number of links, and the total link strength, and negative for the number of clusters, and the intersection $b(i)$ is around 1.0 for the number of terms, the number of links, and the total link strength, and around 0.3 for the number of clusters. Inserting Eq.3 into Eq.2 yields

$$p_{i,\gamma}(n) = B(i,\gamma).n^{\beta(i,\gamma)} \sim B(i,\gamma).n^{[a(i)ln\gamma+b(i)]} = B(i,\gamma).n^{a(i)b(i)}n^{a(i)ln\gamma} \qquad (4)$$

This means that the scaling behavior of our university network is in fact dictated by the occurrence threshold, i.e., by the extent a specific term is present (total number of occurrences) in the set of publications. In the following section we discuss the relevance of our findings for urban scaling research.

### *3.2 Discussion of the findings in the context of city simulation*

Increasing the term occurrence threshold means increasing the minimum number of publications in which a term must occur, which implies that the selection of terms relates more and more to the frequent terms, i.e., to research on current topics. Taking the lowest threshold, $\gamma=1$, i.e., all terms are included, ensures that also less actual topics will be included in the network structure. Particularly less actual topics may represent new, unexpected and original research. In the context of an urban community, this relates to rare abilities and qualities within the population. Shutters et al. (2018) find that the increase in the total network interconnectivity is for an important part driven by the presence of rare, but highly interdependent occupations in a city.

The choice of the occurrence threshold is a choice of diversity. From our observations follows that the occurrence threshold not only determines the total number of terms in the network but also the number of clusters: the higher the occurrence threshold (less term diversity), the less the number of clusters increases with increasing network size, whereas the total link strength of the network increases with a high superlinear scaling exponent. This could indicate that in a growing network the terms with a high occurrence (occurrence higher than 1) lead to a decrease of the complexity of the network measured in number of clusters.

What are the characteristics of our network clusters and how do this relates to cities? We can investigate this by looking at the clusters in Figure 2. The clusters are broad common focus areas that more or less remain the same as the size of the network increases, mostly corresponding to large fields of research. In the Leiden network structure we see large medical and astrophysical clusters, and particularly the medical clusters are characterized by general terms with a high frequency, such as 'behavior', 'survival', 'risk', 'management', 'mortality', 'children'. For lower term occurrence thresholds also more specific, smaller clusters become visible such as dna-expression, aging and dementia, stellar evolution and black holes. This fits well with the idea of the city: there are a more of less fixed number of problems/areas of attention that arise in each city, and some cities have a special clusters such as a port, or a specific industry, or a university. The clusters represent activity centers and we find that they scale sublinearly with increasing ertwork size. We here see a similarity with the work on spatial structure of mobility networks in cities (Louail et al., 2014; 2015; Louf and Bathelemy, 2013; 2014b; 2016). These authors find that the number of hotspots (urban activity centres) scales sublinearly with city population size and they remark that this finding may serve as a guide for constructing a theoretical model.

## 4. Conclusions

In this paper we make an attempt to increase our understanding of the urban scaling phenomenon. In our approach we simulate a city by a set of publications from a large university where the publications take the role of city inhabitants and the terms (keywords) in the publications represent abilities and qualities of the inhabitants. We use in this experiment scientific publications that are connected by co-occurrence of terms to create a network on the basis of these connections, and let this network grow by successively adding publications. We analyze how four important network parameters, namely number of terms, number of clusters, number of links, and total links strength increase with increasing size of the network. This procedure is performed for different values of the term occurrence threshold. This occurrence threshold determines the *diversity of terms*: at a low threshold most or all of the terms are included in the network formation, whereas at a high threshold only the more common, i.e., the more frequently occurring terms are included.

We find a significant power law dependence of these network parameters on network size. We remark that the network grows with the number of publications, but the network itself is a structure of interconnected terms. This is similar to what we want to simulate: the size of cities is determined by the number of inhabitants, whereas the real basis of the urban scaling phenomenon is the network of interactions between all possible abilities and qualities of these inhabitants. The size of a city therefore is a kind of an easy measurable, proxy independent variable. Publications take a similar role in our simulation.

The larger the network (with respect to number publications), the larger the number of clusters, but interestingly this number of clusters is the only network parameter that increases sublinearly (exponent $\beta < 1$) with increasing size of the network. This could be an indication that a growing network tends to decrease complexity by effective self-organization. Professional diversity is an important element for the socioeconomic strength of cities (Bettencourt et al., 2014). In our approach the number of clusters can be interpreted as a measure of complexity within the network. Since the occurrence threshold determines the diversity of terms, we may expect a relation with network complexity, and thus a relation between the occurrence threshold and the number of clusters. This is indeed the case: whereas for the three other network parameters the superlinear scaling exponent increases with increasing occurrence threshold, the number of clusters is the only network parameter of which the sublinear scaling exponent decreases with increasing occurrence threshold. The sublinear scaling of the number of clusters shows a similarity with work on the sublinear scaling of urban activity centers.

If we assume that all terms, but particularly the rare terms, i.e., those with a low occurrence frequency are important for the structure of the network, then the networks with a low occurrence frequency are the most appropriate simulation of the urban complex system and the scaling of this system. Particularly the number of network links and the total network linkage strength are in our opinion the parameters that dominate the scaling phenomenon and can be considered as a simulation of the economic strength of a city, i.e., its gross urban product. The measured power law dependence of these network parameters on network size, and particularly the power law exponents are within the range that is typical for urban scaling.

A low occurrence threshold means a high diversity of terms. We find that highest total link strength for the lowest occurrence threshold. In other words, the more diversity of terms, the stronger the network structure. The phenomenon in cities that may correspond to these observations is that at a high diversity of abilities and capacities, the larger the city the stronger it socio-economically is, and that this socio-economic strength increases slightly superlinear. Another important finding is that the higher the occurrence threshold, the stronger the power law of the number of links and the total link strength increases. In the context of an urban community this finding is intuitively understandable: the more a specific, strongly present ability or quality is present in the urban population (in our simulation higher $\gamma$), the faster the number of links related to that ability will grow for an increasing population. For relatively rare abilities within the urban population (in our simulation low $\gamma$), however important these may be, one can expect that the number of links will grow less fast for an increasing population. We conclude with the key finding if this study: if our simulation model is an appropriate approach to understand urban scaling, the scaling exponent is strongly related to the presence of common but in particular also rare abilities and qualities in the urban population.

### *References*

Altmann EG (2020). Spatial interactions in urban scaling laws. *PLoS ONE* 15(12): e0243390. https://doi.org/10.1371/journal.pone.0243390

Bettencourt LMA, Lobo J (2016). Urban scaling in Europe. *J. R. Soc. Interface* 13: 2016.0005. https://doi.org/10.1098/rsif.2016.0005

Bettencourt LMA (2019). Towards a statistical mechanics of cities. *C. R. Physique* 20(4), 308-318. https://doi.org/10.1016/j.crhy.2019.05.007

Bettencourt LMA, Yang VC, Lobo J, Kempes CP, Rybski D, Hamilton MJ (2020). The interpretation of urban scaling analysis in time. *J. R. Soc. Interface* 17: 20190846. http://dx.doi.org/10.1098/rsif.2019.0846

Braam RR, Moed HF, van Raan AFJ (1991). Mapping of science by combined co-citation and word analysis I: Structural Aspects. *Journal of the American Society for Information Science* 42, 233-251; II: Dynamical Aspects. *Journal of the American Society for Information Science* 42, 252-266. https://doi.org/10.1002/(SICI)1097-4571(199105)42:4<233::AID-ASI1>3.0.CO;2-I

Cebrat C, Sobczyński M (2016). Scaling Laws in City Growth: Setting Limitations with Self-Organizing Maps. *PLoS ONE* 11(12): e0168753. https://doi.org/10.1371/journal.pone.0168753

Cottineau C, Hatna E, Arcaute E, Batty M (2017). Diverse cities or the systematic paradox of urban scaling laws. *Comput Environ Urban Syst* 63:80–94.

Cottineau C, Finance O, Hatna E, Arcaute E, Batty M (2019). Defining urban clusters to detect agglomeration economies. *Environment and Planning B: Urban Analytics and City Science* 46 (6) 1611-1626. https://doi.org/10.1177/2399808318755146

Depersin J, Barthelemy M (2018). From global scaling to the dynamics of individual cities, *Proc Natl Acad Sci USA* 115 (10) 2317–2322. www.pnas.org/cgi/doi/10.1073/pnas.1718690115

Fortunato S, Newman MEJ (2022). 20 years of network community detection. *Nature Phys.* 18, 848–850. https://doi.org/10.1038/s41567-022-01716-7.

Glaeser EL, Scheinkman JA, Shleifer A (1995). Economic growth in a cross-section of cities. *J. Monetary Economics* 36 117-143. https://doi.org/10.1016/0304-3932(95)01206-2

Golosovsky M, Solomon S (2017). Growing complex network of citations of scientific papers: Modeling and measurements. *Phys. Rev.* E 95: 012324. https://doi.org/10.1103/PhysRevE.95.012324.

Jin EM, Girvan M, Newman MEJ (2001). Structure of growing social networks. *Phys. Rev.* E 64: 046132. https://doi.org/10.1103/PhysRevE.64.046132.

Lehmann S, Lautrup B, Jackson AD (2003). Citation networks in high energy physics. *Phys. Rev. E* 68, 026113. https://journals.aps.org/pre/abstract/10.1103/PhysRevE.68.026113

# Supplementary Material

## 1. Network parameters for term occurrence thresholds 1 to 10

For each size of our network structure we calculated with the VoSviewer four parameters: the number of terms involved in the construction of the network, the number of clusters, the number of links, and the total strength of the links. These are our network parameters of which we want to find the dependence on the size of the network (i.e., number of publications with which the network is constructed). We perform these calculations for occurrence thresholds $\gamma=1$ to 10. We also performed our calculations after removing those terms that have a link strength <10 (this value is chosen on the basis of the link strength distribution function, see Figure S3. In Table S1 we show our results for the measured scaling exponents.

Table S1. Scaling exponents for the number of terms (t), the number of clusters (c), the number of links (l), and the total strength of the links (s) for occurrence thresholds $\gamma=1$ to 10. In the left part of the table the exponents are measured with all terms, in the right part the exponents are measured with excluding terms with link strength <10.

| all terms | | | | | terms with links strength $\geq$ 10 | | | | |
|---|---|---|---|---|---|---|---|---|---|
| $\gamma$ | t | c | l | s | $\gamma$ | t | c | l | s |
| 1 | 0.83 | 0.32 | 1.01 | 1.04 | 1 | 0.84 | 0.21 | 1.05 | 1.08 |
| 2 | 1.07 | 0.12 | 1.52 | 1.54 | 2 | 1.17 | 0.05 | 1.57 | 1.59 |
| 3 | 1.24 | 0.05 | 1.81 | 1.84 | 3 | 1.30 | -0.01 | 1.80 | 1.83 |
| 4 | 1.34 | 0.01 | 1.99 | 2.02 | 4 | 1.39 | 0.03 | 1.94 | 1.97 |
| 5 | 1.40 | -0.02 | 2.11 | 2.14 | 5 | 1.49 | 0.02 | 2.09 | 2.12 |
| 6 | 1.51 | 0.13 | 2.27 | 2.28 | 6 | 1.54 | 0.22 | 2.17 | 2.21 |
| 7 | 1.54 | 0.03 | 2.37 | 2.37 | 7 | 1.54 | 0.08 | 2.20 | 2.24 |
| 8 | 1.46 | 0.02 | 2.20 | 2.24 | 8 | 1.56 | 0.00 | 2.24 | 2.27 |
| 9 | 1.54 | 0.01 | 2.33 | 2.36 | 9 | 1.60 | 0.14 | 2.33 | 2.36 |
| 10 | 1.57 | -0.04 | 2.41 | 2.43 | 10 | 1.60 | 0.02 | 2.36 | 2.39 |

## 2. Distribution of link strengths over the terms

There can be a link between any pair of terms. Each link has a specific strength, this strength indicates the number of publications in which two terms occur together. Because a term can be linked to many other terms, we can sum the specific strengths of all links for any term. The VOSviewer allows the calculation of this summed link strength per term. The total of the summed link strengths of all terms is the total link strength of the network as discussed in the main text.

As an example of the distribution of link strengths we show in Figure S1 the rank distribution of the (summed) link strength of the first 100 terms for $\gamma=1$ at $n=3000$ and 6000.

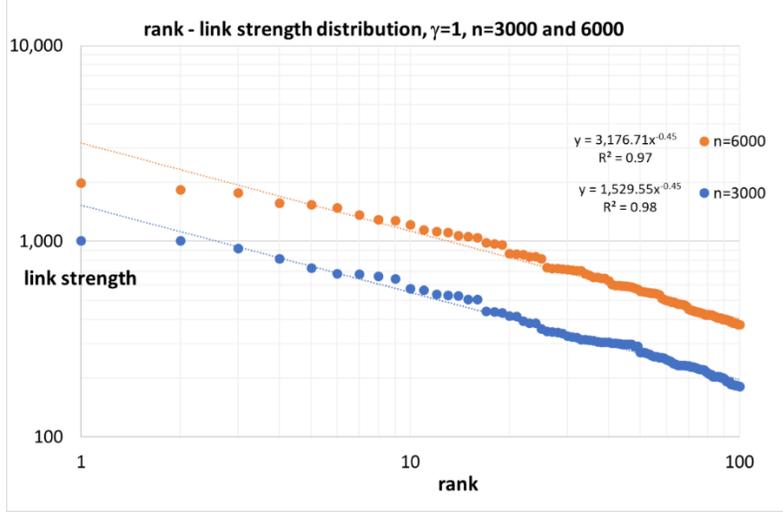

*Figure S1. Rank distribution of the (summed) link strength (S) of the first 100 terms for the network with γ=1 at n=3000 and 6000.*

Remarkably, the distributions are quite similar for both network sizes, particularly the power law dependence. The distribution function has the form

$$S_{n,\gamma}(r) = C(n,\gamma) \cdot r^{-\alpha(n,\gamma)} \qquad (S1)$$

where $S_{N,\gamma}(r)$ is the link strength for a term with rank $r$ in a network with size $n$ and occurrence threshold $\gamma$; $C(n,\gamma)$ is the coefficient of the distribution function and $-\alpha(n,\gamma)$ is the exponent -0.45. This rank distribution resembles a Zipf-distribution (Newman, 2005; Ribeiro et al., 2021). The coefficients however differ, which is obvious as these coefficients are related to the total link strength of the network. We can transform the rank distribution in Eq.S1 into a link strength distribution. To keep eq. S1 simple, we write:

$$S(r) = C \cdot r^{-\alpha} \qquad (S1a)$$

The number of link strengths between $S_1(r_1)$ and $S_2(r_2)$ is obviously $r_2 - r_1$ and thus the number of link strengths $n(S)$ satisfies $n(S)dS = -dr$ so that

$$n(S) = -dr/dS \sim S^{-(1+\alpha)/\alpha} \qquad (S2)$$

which means in our case a link strength distribution of about $S^{-2.54}$ which suggests a scale-free distribution of link strengths over the terms in our networks. This is a similar situation as the work mentioned in the main text on increase of the number of links for a growing network (Li and Lu, 2009; Li, 2009) where also a scale-free network was used.

We see in Figure S1 that the distribution has no typical threshold value within this ranking range. In Figure S2 we show the entire rank-link strength distribution function for all terms for the network with γ=1 at n= 6000. We see a threshold around the summed link strength value 10.

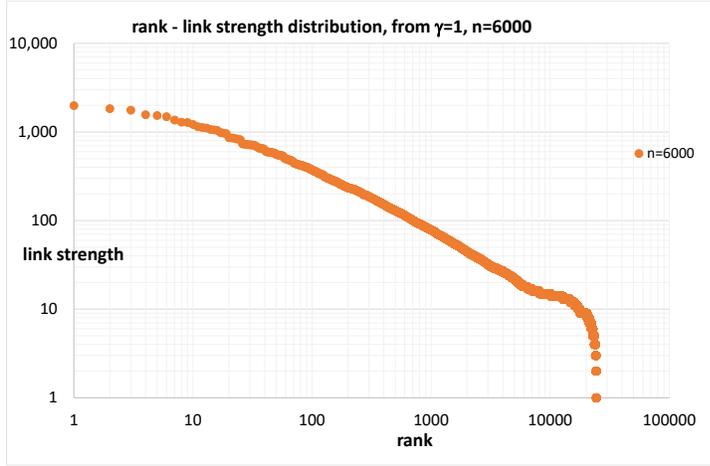

*Figure S2. Rank distribution of the (summed) link strength (S) of all terms for the network with γ=1 at n= 6000.*

### *3. Increase of number of links*

To the best of our knowledge, the only empirical work on the increase of the number of links for a growing network is the study of Li and Lu (2009) and Li (2009). These authors find that the number of links (in their study denoted with $L$, in our study $p_2$) increases with $L \sim m^{1+\theta}$ with $0 \leq \theta \leq 1$. They measure, however, the number of links as s function of the number of nodes $m$, whereas we measure the number of links as a function of network size, i.e., number of publications $n$. With the data in Figure 3 in the main text it is easy to find out how the number of links increases with the number of terms, which are the nodes in our network. The mathematical derivation of this relation is given here below. We find for $\gamma =1$ a power law dependence with exponent 1.21. Li and Lu find for $1+\theta$ values between 1.12 and 1.31 which is close to our values.

We specify Eq.2 in the main text for number of links $p_2$, and the number of terms $p_3$:

$$p_{2,\gamma}(n) = B(2,\gamma) \cdot n^{\beta(2,\gamma)} \tag{1a}$$

$$p_{3,\gamma}(n) = B(3,\gamma) \cdot n^{\beta(3,\gamma)} \tag{1b}$$

Exponentiating both sides of Eq.1b with $\beta(2,\gamma)/\beta(3,\gamma)$ yields

$$p_{3,\gamma}(n)^{\frac{\beta(2,\gamma)}{\beta(3,\gamma)}} = B(3,\gamma)^{\frac{\beta(2,\gamma)}{\beta(3,\gamma)}} \cdot n^{\beta(2,\gamma)} \tag{1c}$$

hence

$$n^{\beta(2,\gamma)} = p_{3,\gamma}(n)^{\frac{\beta(2,\gamma)}{\beta(3,\gamma)}} / B(3,\gamma)^{\frac{\beta(2,\gamma)}{\beta(3,\gamma)}} \tag{1d}$$

Inserting the right side of Eq.1d into Eq.1a gives

$$p_{2,\gamma}(n) = [B(2,\gamma)/B(3,\gamma)^{\frac{\beta(2,\gamma)}{\beta(3,\gamma)}}].p_{3,\gamma}(n)^{\frac{\beta(2,\gamma)}{\beta(3,\gamma)}} \quad (1e)$$

which means that the number of links (at a given number of publications $n$) is related to the number of terms by a power law with an exponent equal to the ratio $\frac{\beta(2,\gamma)}{\beta(3,\gamma)}$.

## *4. Influence of term occurrence and of occurrence threshold*

It is interesting to find out how specific occurrence values play a role in the structure of the network. In Figure S3 we present the relation between the number of terms and the occurrence of a term, and the relation between the occurrence of a term and its link strength in the network. We see that the number of terms decreases as a function of occurrence with a power law exponent -2.33. On the other hand, the higher the occurrence of a term, the larger its link strength in the network, with a linear relation. Thus, the higher the occurrence of a term, the lower the number of these terms, but the larger its contribution to the total link strength of the network.

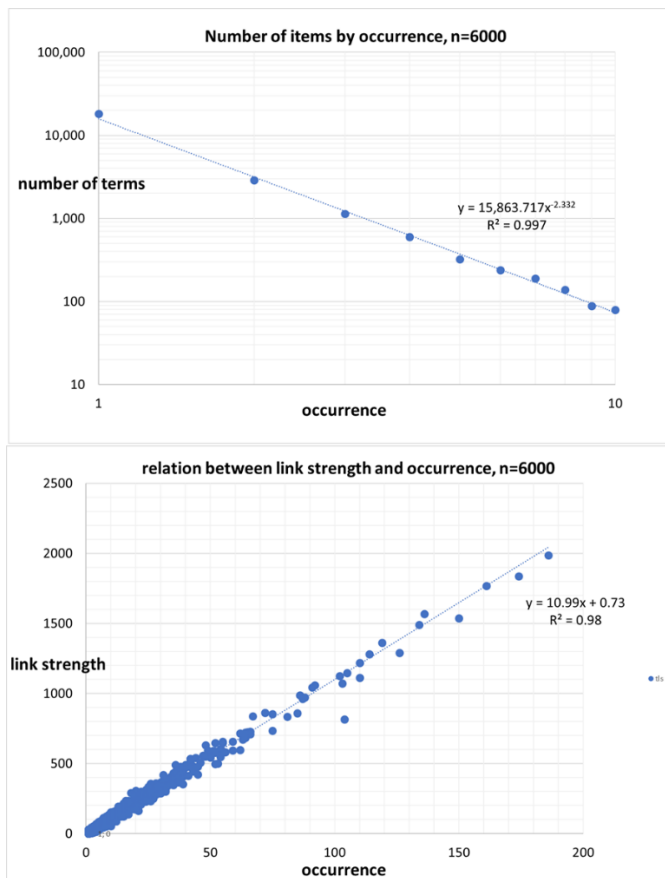

*Figure S3. The relation between number of terms and occurrence of a term (upper panel), and the relation between occurrence of a term and its link strength in the network (lower panel).*